\documentclass[prl,aps,twocolumn,showpacs,preprintnumbers,amsmath,amssymb]
{revtex4}

\usepackage{graphicx}
\usepackage[dvips]{color}

\begin{document}


\title{Stability of superfluid and supersolid phases of dipolar bosons in optical lattices}
\author{Ippei Danshita$^{1}$}
\author{Carlos A. R. S\'a de Melo$^{2}$}
\affiliation{
{$^1$Department of Physics, Waseda University, Shinjuku-ku, Tokyo 169-8555, Japan}
\\
{$^2$School of Physics, Georgia Institute of Technology, Atlanta, Georgia 30332, USA}
}

\date{\today}

\begin{abstract}
We perform a stability analysis of 
superfluid (SF) and supersolid (SS) phases of polarized dipolar bosons 
in two-dimensional optical lattices at high filling factors and zero temperature, 
and obtain the phase boundaries between SF, checkerboard SS (CSS),
striped SS (SSS), and collapse. 
We show that the phase diagram can be explored through the application of 
an external field and the tuning of its direction with respect to 
the optical lattice plane. 
In particular, we find a transition between the CSS and SSS phases.
\end{abstract}

\pacs{03.75.Hh, 03.75.Lm, 05.30.Jp}

\keywords{optical lattice, Bose-Einstein condensation, 
superfluid, supersolid, dipolar boson, dynamical instability}
\maketitle
Since the realization of the superfluid-Mott insulator transition 
of ultracold Bose gases confined to optical lattices, 
ultracold atoms have become the playground
for the realization of various quantum phases studied 
in condensed matter physics~\cite{bloch-2005}.
The unprecedented control of the lattice depth, dimensionality, geometry, and 
filling factor has allowed for the exploration of a variety of effects, 
including the experimental observation of a dipolar condensate 
of $^{52}{\rm Cr}$ atoms~\cite{griesmaier-2005, stuhler-2005,lahaye-2007}
and the production of ultracold heteronuclear molecules~\cite{ospelkaus-2006}.

The problem of interacting dipolar bosons is important not only for magnetic 
dipolar atoms, but also for heteronuclear molecules and Rydberg atoms,
which can have potentially large electric dipole moments. 
Thus far, dipolar superfluids (SF) have been found only for magnetic 
dipolar atoms in optical traps~\cite{griesmaier-2005, 
stuhler-2005,lahaye-2007}, but there are still several phases that can be 
pursued experimentally, including dipolar supersolids (SS), 
which are characterized by the simultaneous existence of crystalline and
SF orders.
The possibility of SS phases first emerged in the context of 
solid $^{4} {\rm He}$~\cite{andreev-1969,chester-1970,leggett-1970}. 
More recently, there have been 
experimental reports that the theoretically predicted~\cite{leggett-1970} 
non-classical rotational inertia was found~\cite{kim-2004} 
in solid $^{4}{\rm He}$.
Although the existence of a SS still remains a controversial issue 
in the condensed matter literature~\cite{sasaki-2006},  
its existence may be easier to verify in the context of Bose gases 
in optical lattices.

In this manuscript, we analyze SF and SS phases of 
ultracold dipolar Bose gases loaded into two-dimensional (2D) optical lattices, 
and focus on the region of high filling factors.
We show that using an external field the sign and magnitude of dipole 
interactions can be controlled leading to a variety of different phases. 
The phases described include SF, striped SS (SSS), 
checkerboard SS (CSS), and collapse, 
for which we analyze experimentally relevant quantities such as 
the excitation spectra.

Our work on ultracold dipolar bosons distinguishes itself from recent work 
on this topic in several ways. 
First, our discussion is analytical 
in contrast to numerical work 
based on the Gutzwiller projection techniques~\cite{goral-2002,
menotti-2007, yi-2007} or quantum Monte Carlo (QMC)
methods~\cite{QMCs-1990s,sengupta-2005,QMCs-hardcore}.
Second, while some work~\cite{menotti-2007, kovrizhin-2005} was confined 
to purely repulsive dipolar interactions, we allow for the competition 
between attractive and repulsive dipolar interactions by changing the 
direction of an external field with respect to the 2D lattice plane.
Third, we explore the phase diagram for a wider range of interactions 
than previously investigated~\cite{goral-2002,menotti-2007,yi-2007, kovrizhin-2005} 
and find that the quantum phase of the system can be switched between
CSS and SSS as the direction of the dipole is changed.
The transition between these SS phases is remarkable in the sense
that it is a structural phase transition between two types of SS,
where the symmetry of the crystalline order changes.
Lastly, we go beyond mean-field descriptions by including fluctuation effects 
and performing a stability analysis.

\begin{figure}[b]
\includegraphics[scale=0.26]{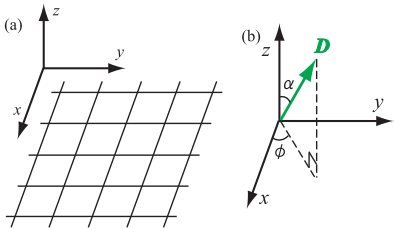}
\caption{\label{fig:dipolar}
(color online) (a) Schematic picture of a 2D lattice system, where a square 
represents a site. (b) The dipole vector is shown in the spherical polar 
coordinates.
}
\end{figure}

To study bosons with dipole-dipole interactions in 2D optical lattices, 
we use the dipolar-Bose-Hubbard model
\begin{eqnarray}
H\!=\! 
-J\!\sum_{\langle jl \rangle}
(b^{\dagger}_{j} b_{l}\!+\!{\rm h.c.})
\!+\!\frac{U}{2}\! \sum_{j}\!
n_{j} (n_{j}\!-\!1)
\!+\!\!\sum_{ j < l }\!V_{jl}
n_{j}n_{l},
\label{eq:hamiltonian}
\end{eqnarray}
where $b^{\dagger}_{j}$ is the boson creation operator at site $j$,
$n_j=b^{\dagger}_{j} b_{j}$, and $J$ is the hopping. 
$\langle jl \rangle$ represents nearest-neighbor (NN) pairs of lattice sites.
The onsite interaction $U=U_s + U_d$ consists of the contributions 
from both the s-wave scattering
$U_s=4\pi\hbar^2 a_s/m \int d{\bf r} |w({\bf r})|^4$
and the onsite dipole-dipole interaction 
$U_d=(2\pi)^{3/2}\int d{\bf k} \tilde{n}^2({\bf k})\tilde{V}_d({\bf k})$~\cite{menotti-2007, menotti-2008},
where $a_s$ is the s-wave scattering length, $w({\bf r})$ is the Wannier function 
with respect to the underlying lattice potential, and $\tilde{n}({\bf k})$ and 
$\tilde{V}_d({\bf k})$ are the Fourier transform of the density $|w({\bf r})|^2$ and 
the dipole-dipole potential.
The long-range part of the dipole-dipole interaction is well-approximated as 
$
V_{jl}=D^2(1-3{\rm cos}^2\theta_{jl})
                |{\bf r}_j-{\bf r}_l|^{-3}
$
where $D$ is the dipole moment and $\theta_{jl}$ is the angle between 
the dipole direction and ${\bf r}_j-{\bf r}_l$.
Here, ${\bf r}_j=(j_x a,j_y a)$ is a lattice vector, 
where $j_x$ and $j_y$ are integers and $a$ is the lattice spacing.
A schematic picture of the system is shown in Fig.~\ref{fig:dipolar}.

%
%
Since we are interested only in
SF and SS phases, we restrict ourselves to Bose-condensed 
solutions of Eq.~(\ref{eq:hamiltonian}).
Minimizing the quantum action 
$S = i\hbar \sum_j b_j^{\dagger} \partial_t b_j - H$
with respect to the saddle-point field $\widetilde \Psi_j (t)$ obtained 
from the transformation
$b_j (t) \to \Psi_j (t) =  \widetilde \Psi_{j} (t) + \delta \Psi_j (t)$, 
and neglecting the fluctuations $\delta \Psi_j (t)$ lead to the 
time-dependent Gross-Pitaevskii (GP) equation
%
$
i\hbar \partial_t \widetilde\Psi_{j} = 
-J\sum_{\langle l \rangle} \widetilde\Psi_{l}
 + (
U |\widetilde\Psi_{j}|^2+\sum_{l\neq j}V_{jl}|\widetilde\Psi_{l}|^2
)\widetilde\Psi_{j}
$.
%
The description above is justified when 
the healing length $\xi = \hbar / (m^{\ast}c)$ is larger than the interatomic 
distance $l=a/\sqrt{\nu}$~\cite{pitaevskii-2003}, where $c$ is the sound
velocity, $m^{\ast}$ is the effective mass, and $\nu$ is the filling factor.
This condition is satisfied when $J\gg {\rm max}(U,|V_{i,j}|)$.
In addition, as discussed later, the emergence of SS phases 
requires $J < \nu U$, which can be satisfied when $\nu \gg 1$.
Thus, our approach is valid for large hopping and large
filling factor.
Notice that we are not investigating incompressible solid phases, 
such as Mott-insulators and density wave solids, because they can not be described 
within this GP approximation.

Writing $\widetilde\Psi_{j} (t) = \Phi_j e^{-i\mu t/\hbar} $ requires 
the static part of the condensate $\Phi_{j}$ to satisfy 
\begin{eqnarray}
-J\sum_{\langle l \rangle}\Phi_l
+\bigl(
U |\Phi_j|^2+\sum_{l\neq j}V_{j,l}|\Phi_l|^2
\bigr)\Phi_j=\mu\Phi_j,
\label{eq:sGPE}
\end{eqnarray}
where $\mu$ is the chemical potential.
The normalization condition is 
$
M^{-1} \sum_{j}|\Phi_j|^2=\nu,
$
where $M$ is the total number of sites; and the energy of the condensate is 
\begin{eqnarray}
E \! = \!
-J\!\sum_{\langle jl \rangle}
\! (\Phi^{\ast}_{j} \Phi_{l}\!+\!{\rm c.c.})
\! + \! \frac{U}{2} \! \sum_{j}
\! |\Phi_j|^4
\! + \!\! \sum_{j < l }\! V_{jl}
|\Phi_{j}|^2 |\Phi_{l}|^2.
\label{eq:totalenergy}
\end{eqnarray}
%

A standard procedure to obtain phase diagrams is the comparison
of energies for various phases as done in mean-field studies based 
on the Gutzwiller projection technique
~\cite{goral-2002, menotti-2007, yi-2007}. 
It is also important to study the stability 
of each phase in order to construct the correct phase diagram. 
For this purpose, we perform a stability analysis by taking into account 
fluctuations 
%
%
$
\delta\Psi_j = 
e^{-i\mu t/\hbar}
\bigl( 
u_j e^{-i\varepsilon t/\hbar} 
-v_j^{\ast} e^{i\varepsilon^{\ast} t/\hbar} 
\bigr)
$
beyond the saddle-point $\widetilde\Psi_{j}$, 
which lead to the eigenvalue equations
\begin{eqnarray}
&&\!\!\!\!\!\!\!\!
-J\sum_{\langle l \rangle}u_l
+\biggl(
2U |\Phi_j|^2+\sum_{l\neq j}V_{j,l}|\Phi_l|^2-\mu
\biggr)u_j
\nonumber\\
&&\!\!\!\!\!\!\!\!+
\Phi_j\sum_{l\neq j}V_{j,l}\Phi_l^{\ast}u_l
\! - \! U\Phi_j^2 v_j -\Phi_j\sum_{l \neq j}V_{j,l}\Phi_l v_l
\! =\! \varepsilon u_j,
\label{eq:BdGE1}
\\
&&\!\!\!\!\!\!
-J\sum_{\langle l \rangle}v_l
+\biggl(
2U |\Phi_j|^2+\sum_{l\neq j} V_{j,l}|\Phi_l|^2-\mu
\biggr)v_j
\nonumber\\
&&\!\!\!\!\!\!\!\!\!\!\!\!\!\! +
\Phi_j^{\ast} \! \sum_{l\neq j} \! V_{j,l}\Phi_l v_l 
\!-\!U(\Phi_j^{\ast})^2 u_j 
\!-\!\Phi_j^{\ast} \! \sum_{l \neq j} \! V_{j,l}\Phi_l^{\ast} u_l
\!=\!
-\varepsilon v_j.
\label{eq:BdGE2}
\end{eqnarray}
Here, $\varepsilon$ is the energy and $(u_j,v_j)$ is the amplitude of
the collective mode of the condensate.
The appearance of collective modes with complex frequencies signals 
exponential growth of fluctuations in time, and thus the existence of
dynamical instabilities (DI). 

Since the dipolar interaction decays over distance, the ratios between
the NN interactions $V_x=(1-3\sin^2\alpha\cos^2\phi)D^2/a^3$, 
$V_y=(1-3\sin^2\alpha\sin^2\phi)D^2/a^3$ (along the $x$ and $y$ directions) 
and the onsite interaction $U$ dictates the basic physics.
Here, $\alpha$ and $\phi$ are the elevation and azimuthal angles of
the direction of the polarization with respect to the 2D lattice plane
as shown in Fig.~\ref{fig:dipolar}(b).
These ratios are experimentally controllable, e.g., by changing the $s$-wave 
scattering length via a Feshbach resonance, 
as demonstrated with $^{52}$Cr atoms~\cite{lahaye-2007}. 
Moreover, the ratio $\gamma = V_y/V_x$ is also controllable by changing 
the polarization direction through the application of an external 
field.  
For instance, when $\phi = \pi/2$ and $\alpha$ varies, the relevant ratio 
becomes $\gamma = 1 - 3\,{\rm sin}^2\alpha$.

\begin{figure}[tb]
\includegraphics[scale=0.4]{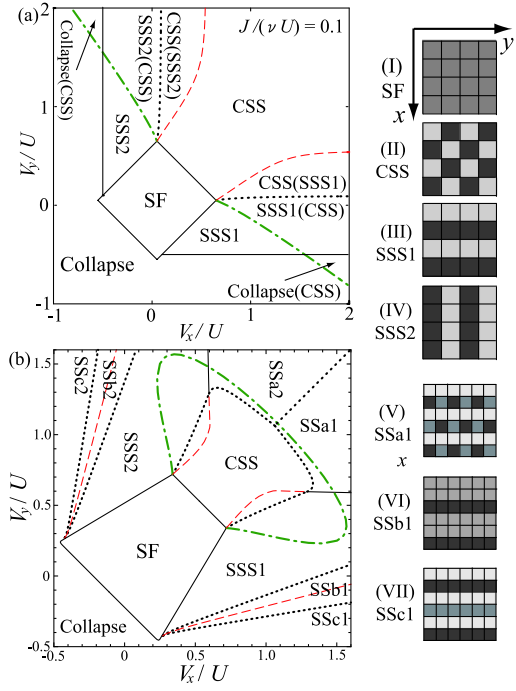}
\caption{\label{fig:phasediagram}
(color online)
Left: Phase diagrams in the $(V_x/U,V_y/U)$ plane for $J/(\nu U)=0.1$. 
While only the NN interactions are considered in (a), 
the full dipolar interactions are included in (b).
The solid (dotted) lines represent phase boundaries  
for continuous (discontinuous) transitions.
The dashed-dotted green lines and the dashed red lines separate the stable and unstable regions 
of the CSS state and those of the SSS. 
Right: Schematic pictures of SF (I), CSS (II), ${\rm SSS1}$ (III), ${\rm SSS2}$ (IV),
SSa1 (V), SSb1 (VI), and SSc1 (VII).
}
\end{figure}

To gain analytical insight, we consider first onsite and NN interactions.
In this case, there are four possible pure condensate phases.
When the absolute values of $V_x$ and $V_y$ are much smaller than
$U$, the system is in a SF phase.
When the NN interactions are almost isotropic and strongly repulsive, diagonal
crystalline order develops and a CSS phase emerges. 
When the NN interaction is strongly anisotropic, e.g. $V_x= - V_y$, 
a SSS phase is favored.
Schematic pictures of these phases are shown in 
Figs.~\ref{fig:phasediagram}(I)-(IV).
The phase boundaries of SF, CSS, and SSS can be determined by solving
Eq.~(\ref{eq:sGPE}) and obtaining the ground state energy given in 
Eq.~(\ref{eq:totalenergy}).
In addition, strongly attractive NN interactions result in the collapse
of the condensate, which is characterized by a DI
occurring in the long-wavelength phonon mode.
In Fig.~\ref{fig:phasediagram}, the solid and dotted lines indicate the phase boundaries 
in the $(V_x/U,V_y/U)$ plane for $J/(\nu U)=0.1$.

%
%

%
%

Let us first identify the SF region.
In the SF phase, the condensate wave function,
$
\Phi_j=\sqrt{\nu},
$
is uniform and the chemical potential is 
$
\mu_{\rm SF} = -4J+\nu(U+2V_x +2V_y).
$
From Eq.~(\ref{eq:totalenergy}), we obtain 
the energy
$E_{\rm SF}/M=-4J\nu+\nu^2(U/2+V_x + V_y)$, 
and from Eqs.~(\ref{eq:BdGE1}) and (\ref{eq:BdGE2})
we obtain the excitation spectrum
$
\varepsilon_{\rm SF}({\bf q}) = \sqrt{ \tilde{\epsilon}({\bf q})\left[
\tilde{\epsilon}({\bf q})+2\nu(U+2V({\bf q}))\right]},
$
where ${\bf q}$ is the quasimomentum of the collective mode,
$
\tilde{\epsilon}({\bf q})=
4J\left({\rm sin}^2(q_x a/2)+{\rm sin}^2(q_y a/2)\right)$,
and
$
V({\bf q})=V_x {\rm cos}(q_x a) + V_y {\rm cos}(q_y a).
$

Since $\varepsilon_{\rm SF}({\bf q})= c_{\rm SF}\hbar q$ for $qa\ll 1$,
where $c_{\rm SF} = \sqrt{2\nu(U + 2V_x + 2V_y)J} a/\hbar$ is the sound
velocity for the SF phase, long-wavelength phonons cause a DI
when $U + 2V_x + 2V_y < 0$ for any non-zero value of $J$. 
The sound velocity $c$ is directly related to the compressibility 
$\kappa$ via  $\kappa^{-1} = m^{\ast}c^2$,
which clearly shows that the compressibility becomes negative 
at this DI leading to the collapse of the condensate 
~\cite{rychtarik-2004}.
On the other hand, since
$\varepsilon_{\rm SF}({\bf Q}_{0}) = \sqrt{8J + 2\nu (U-2V_x-2V_y)}$, 
where ${\bf Q}_{0}\equiv (\pi/a,\pi/a)$, the collective modes in the vicinity 
of ${\bf Q}_{0}$ cause a DI when $2V_x + 2V_y - U - 4J/\nu>0$,
signaling a transition to the CSS phase~\cite{kovrizhin-2005}, where the mode 
with ${\bf q}={\bf Q}_{0}$ is amplified and its interference with the condensate 
in the zero momentum state creates the checkerboard density wave order.
Similarly, we find that the collective modes in the vicinity of
${\bf q}={\bf Q}_{1}\equiv (\pi/a,0)$ or 
${\bf Q}_{2}\equiv (0,\pi/a)$ cause a DI
and signal transitions to the SSS phases, when 
$2V_x-2V_y-U-2J/\nu>0$ or $2V_y-2V_x-U-2J/\nu>0$, respectively.
Thus, the SF phase corresponds to the square region surrounded by the lines 
$2V_x + 2V_y + U = 0$, $2V_x+2V_y-U-4J/\nu=0$, $2V_x-2V_y-U-2J/\nu=0$ and 
$2V_y-2V_x-U-2J/\nu=0$ as shown in Fig.~\ref{fig:phasediagram}(a).


Since the phase diagram is symmetric with respect to the line $V_x = V_y$, 
we focus on the region $V_x > V_y$ henceforth. The condensate wave functions 
for the CSS and ${\rm SSS1}$ states are 
$
\Phi_{j}^{\rm CSS}=\sqrt{\nu_{\rm c_0}}+\sqrt{\nu_{\rm c}}
e^{i{\bf Q}_0 \cdot{\bf r}_j}
$ 
and
$
\Phi_{j}^{\rm SSS1}=\sqrt{\nu_{\rm s_0}}+\sqrt{\nu_{\rm s}}
e^{i{\bf Q}_1 \cdot{\bf r}_j}
.
$
Here, $\nu_{\rm c}$ ($\nu_{\rm s}$) is the density of atoms condensed 
in the state with quasimomentum ${\bf Q}_{0}$ (${\bf Q}_{1}$), 
corresponding to the order parameter of CSS (${\rm SSS1}$). 
These states are regarded as SS phases in the sense that each of 
them possesses off-diagonal long-range order through the 
condensate, and diagonal long-range order via a density wave. 
From Eq.~(\ref{eq:sGPE}), for the CSS phase we obtain 
$\mu_{\rm CSS} = 2\nu U$,
$\nu_{\rm c_0}=\nu-\nu_{\rm c}$, and
$\nu_{\rm c}=\nu/2-2J/(2V_x + 2V_y-U)$.
Similarly for ${\rm SSS1}$, we obtain
$\mu_{\rm SSS1} = -2J+2\nu (U + 2V_y)$,
$\nu_{\rm s_0} = \nu-\nu_{\rm s}$, and
$\nu_{\rm s}=\nu/2-J/(2V_x-2V_y-U)$.
Since $\nu_{\rm c}$ $(\nu_{\rm s})$ vanishes 
at the phase boundary between SF and CSS (${\rm SSS1}$), 
the phase transitions are continuous according to Landau's classification
(or second-order according to Ehrenfest's).

From Eq.~(\ref{eq:totalenergy}), the energies of the condensate for the CSS
and the ${\rm SSS1}$ are expressed as
%
$
E_{\rm CSS}/M=-8J^2/(2V_x + 2V_y -U)+\nu^2 U,
$
and
$
E_{\rm SSS1}/M=-2J\nu-2J^2/(2V_x - 2V_y -U) + \nu^2(2V_y + U).
$
%
The condition, $E_{\rm CSS} = E_{\rm SSS1}$, determines the phase boundary
between CSS and ${\rm SSS1}$ as shown in Fig.~\ref{fig:phasediagram}(a).
Both $\nu_{\rm c}$ in CSS and $\nu_{\rm s}$ 
in ${\rm SSS1}$ are finite at the phase boundary.
This fact indicates that the phase transition between CSS and 
${\rm SSS1}$ is discontinuous (first-order).
A discontinuous phase transition is characterized by hysteresis, 
such that the critical point $V_y^{\rm d_1}$ of the transition from  
CSS to ${\rm SSS1}$ disagrees with the critical point $V_y^{\rm d_2}$ of transition from 
${\rm SSS1}$ to CSS due to the presence of more than one meta-stable state.
To determine $V_y^{\rm d_1}$ and $V_y^{\rm d_2}$, we calculate 
the excitation spectra $\varepsilon_{\rm CSS}^{\pm}({\bf q})$ and
$\varepsilon_{\rm SSS1}^{\pm}({\bf q})$ for CSS and ${\rm SSS}1$, 
where the plus (minus) sign corresponds to a gapful (gapless) mode at 
low momenta.

\begin{figure}[t]
\includegraphics[scale=0.45]{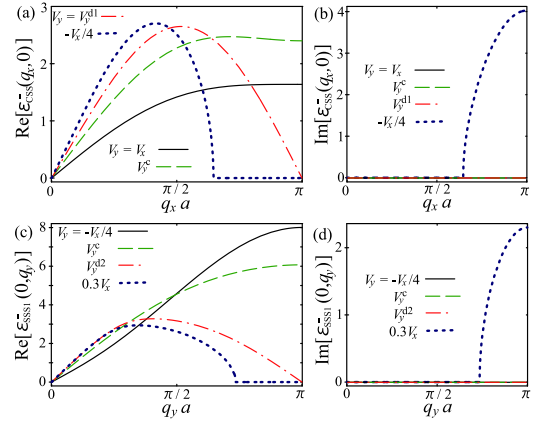}
\caption{\label{fig:excitation}
(color online)
Excitation spectra $\varepsilon_{\rm CSS}^{-}(q_x,q_y=0)$ and
$\varepsilon_{\rm SSS1}^{-}(q_x=0,q_y)$ for $J/(\nu U)=0.1$, 
and $V_x=U$. 
(a) The real part and (b) the imaginary parts of 
$\varepsilon_{\rm CSS}^{-}(q_x,q_y=0)$.
(c) The real part and (d) the imaginary parts of 
$\varepsilon_{\rm SSS1}^{-}(q_x=0,q_y)$.
}
\end{figure}

In Figs.~\ref{fig:excitation}(a) and (b),
we show $\varepsilon_{\rm CSS}^{-}(q_x,q_y=0)$ for 
varying $V_y$, but fixed $V_x = U$. 
As $V_y$ decreases, a roton-like minimum is formed at ${\bf q}={\bf Q}_{1}$
and reaches zero at $V_y = V_y^{\rm d_1}$.
As $V_y$ is decreased further, the imaginary part of 
$\varepsilon_{\rm CSS}^{-}({\bf Q}_{1})$ grows, thus revealing the transition 
to the ${\rm SSS1}$ phase.
The condition $\varepsilon_{\rm CSS}^{-}({\bf Q}_{1})= 0$ then 
gives $V_y^{\rm d_1}$ as shown by the dashed-dotted green line 
in Fig.~\ref{fig:phasediagram}(a).
In Figs.~\ref{fig:excitation}(c) and (d), 
we show $\varepsilon_{\rm SSS1}^{-}(q_x = 0,q_y)$ for
varying $V_y$, but fixed $V_x = U$.
The condition $\varepsilon_{\rm SSS1}^{-}({\bf Q}_{2})= 0$ gives 
$V_y^{\rm d_2}$ as shown by the dashed red line
in Fig.~\ref{fig:phasediagram}(a).

To complete our analysis of the phase diagram, 
we locate now the boundary between phases ${\rm SSS1}$ and collapse.
When 
$qa \ll 1$,
\begin{eqnarray}
\varepsilon_{\rm SSS1}^{-}({\bf q})
\!\simeq\!
\sqrt{2V_y \!+\! U}\!\left(\!\frac{8J^2(q_x a)^2}{2V_x \! -\! 2V_y \!-\! U}
\!+\! 4\nu J (q_y a)^2\!\right)^{1/2}\!, 
\label{eq:collapse}
\end{eqnarray}
and the DI leads to the collapse 
when $2 V_y + U <0$ (see Fig.~\ref{fig:phasediagram}(a)).

Next we numerically calculate the energy and perform linear stability analyses
for the full dipolar interactions.
We choose a specific case of $\phi=0$ or $\pi/2$, where the dipolar interactions 
can be written as $V_{jl}=(i_x^2 V_x + i_y^2 V_y)/(i_x^2+i_y^2)^{5/2}$. $i_x$ 
and $i_y$ are integers that satisfy ${\bf r}_j - {\bf r}_l = (i_x a, i_y a)$.
In this case, we obtain the phase diagram in the $(V_x/U,V_y/U)$-plane for 
$J/(\nu U) = 0.1$ as shown in Fig.~\ref{fig:phasediagram}(b)~\cite{footnote}.
As well as in the phase diagram with only the NN interactions, 
there remain the CSS and SSS regions, which we have confirmed to be
stable.
CSS shares phase boundaries with SSS and the transition 
between these two phases is discontinuous (first-order).
However, unlike the case with only the NN interactions, there are
additional SS phases with different density wave orders, such as SSa,
SSb, and SSc (see (V)-(VII) of Fig.~\ref{fig:phasediagram}).
While SSb has two-sublattice density modulation as well as CSS and SSS do,
SSa and SSc have three-sublattice modulation.
There can be other SS phases with more than three-sublattice modulation,
which shares boundary with SSa and SSc, but here we do not determine 
the location of these phases because we are mainly interested
in the SF, CSS, and SSS phases.

In phenomenological classical models involving 
dipolar interactions for 2D continuum 
systems~\cite{kivelson,ng,spivak} it has been conjectured that 
superstructures or micro-emulsion phases exist between two stable phases
in a region of the temperature versus density phase diagram 
where phase separation overrides an otherwise expected first order
phase transition.
A necessary condition for
the emergence of such micro-emulsions at zero temperature is 
the negative compressibility leading to phase separation. 
For a first order transition between CSS and SSS 
described within our model and approach, we do not find a negative compressibility, 
namely the pre-condition for phase separation, in the transition region.
In contrast, recent QMC simulations by Pollet {\it et al.} have 
shown preliminary evidence of micro-emulsion phases
in systems of hardcore dipolar bosons
in a 2D triangular lattice~\cite{pollet}.
However, QMC simulations of larger system sizes 
are necessary to make a conclusive statement about the presence of 
the micro-emulsion phases.

%
%

In conclusion, we have studied stability of quantum phases of
dipolar bosons in two-dimensional optical lattices. We have shown 
that mean-field theories alone fail to describe correctly the
phase boundaries, and that a stability analysis is absolutely
necessary.  In addition to superfluid and collapsed phases, we 
have shown that striped and checkerboard supersolids can exist, and compete
with each other due to the anisotropy of dipolar interactions 
as controlled by an external field.

%
%
We thank D. Yamamoto, L. Pollet and R. Scalletar for useful discussions,
and the JQI for their hospitality. I. D. is supported by a 
Grant-in-Aid from JSPS, and C. SdM thanks NSF (DMR-0709584) 
and ARO (W911NF-09-1-0220) for support. 

%
%

\end{document}